\documentclass[aps,prd,preprint,superscriptaddress,showpacs,floatfix,nobibnotes]{revtex4-1}

\usepackage{latexsym}
\usepackage{amsmath}
\usepackage{amssymb}
\usepackage{graphicx}

\usepackage{bm}
\usepackage{epsfig}
\usepackage{subfigure}
\newcommand{\bea}{\begin{eqnarray}}
\newcommand{\eea}{\end{eqnarray}}

\begin{document}

\title{The 2PI effective theory at next-to-leading order using the functional renormalization group}

\author{M.E. Carrington}
\email[]{carrington@brandonu.ca} 
\affiliation{Department of Physics, Brandon University, Brandon, Manitoba, R7A 6A9 Canada}\affiliation{Winnipeg Institute for Theoretical Physics, Winnipeg, Manitoba}
 
\author{S.A. Friesen}
\email[]{friesenseth@gmail.com} 
\affiliation{Department of Physics, Brandon University, Brandon, Manitoba, R7A 6A9 Canada}

\author{B.A. Meggison}
\email[]{brett.meggison@gmail.com} 
\affiliation{Department of Physics, Brandon University, Brandon, Manitoba, R7A 6A9 Canada}\affiliation{Winnipeg Institute for Theoretical Physics, Winnipeg, Manitoba}
\affiliation{Department of Physics, University of Manitoba, Winnipeg, Manitoba R3T 2N2}

\author{C.D. Phillips}
\email[]{christopherdphillips7@gmail.com} 
\affiliation{Department of Physics, Brandon University, Brandon, Manitoba, R7A 6A9 Canada}

\author{D. Pickering}
\email[]{pickering@brandonu.ca} 
\affiliation{Department of Mathematics, Brandon University, Brandon, Manitoba, R7A 6A9 Canada}

\author{K. Sohrabi}
\email[]{sohrabik@brandonu.ca} 
\affiliation{Department of Physics, Brandon University, Brandon, Manitoba, R7A 6A9 Canada}

\date{\today}

\begin{abstract}

We consider a symmetric scalar theory with quartic coupling in 4-dimensions.
We show that the 4 loop 2PI calculation can be done using a renormalization group method. 
The calculation involves one bare coupling constant which is introduced at the level of the Lagrangian and is therefore conceptually simpler than a standard 2PI calculation, which requires multiple counterterms. 
We explain how our method can be used to do the corresponding calculation at the 4PI level, which can not be done using any known method by introducing counterterms. 
\end{abstract}


\normalsize
\maketitle

\normalsize

\section{Introduction}
\label{section-introduction}

There are many interesting systems for which there is no small expansion parameter that could be used to implement a perturbative approach, and for this reason much work has been done in recent years on the development non-perturbative methods. Lattice calculations are valuable in situations where the underlying microscopic theory is known, but issues with the continuum and finite volume limit arise. 
Various forms of reorganized/improved hard-thermal-loop resummations have been formulated and applied to the calculation of thermodynamic quantities \cite{htl-extensions1,htl-extensions2,htl-extensions3,htl-extensions4,htl-extensions5,htl-extensions6,htl-extensions7,htl-extensions8,htl-extensions9}.
Schwinger-Dyson equations are another popular and familar approach to non-perturbative problems (for a classic introduction see \cite{sch-dyson1}, a more recent review can be found in \cite{sch-dyson2}). One significant issue with the Schwinger-Dyson approach is that the hierarchy of coupled equations must be truncated by introducing some external prescription. Various methods to construct a truncation that preserves gauge invariance have been proposed \cite{pinch,binosi2,Pennington2009}.

Another powerful technique is the renormalization  group \cite{wetterich}, which is traditionally used to study systems where scale-dependent behaviour is important. Its functional formulation can be cast into the form of an exact flow equation for the scale-dependent effective action. A hierarchy of coupled flow equations for the $n$-point functions of the theory can be obtained from the action flow equation, but this hierarchy must again be truncated using some prescription \cite{BMW1,BMW2}. Some useful reviews include \cite{rg-review1,rg-review2,rg-review3,rg-review4,rg-review5,rg-review6,rg-review7}.

The $n$-particle-irreducible effective action is another method to include non-perturbative effects. In the context of non-relativistic statistical mechanics, the original formalism can be found in Refs. \cite{Yang, Luttinger, Martin}. In its modern form, the method involves writing the action as a functional of dressed vertex functions, which are determined self consistently using the variational principle \cite{Jackiw1974,Norton1975}. 
The technique has been used to study the thermodynamics of quantum fields \cite{Blaizot1999,Berges2005a,meggison}, transport coefficients \cite{Aarts2004,Carrington2006,Carrington-transport-3pi,carrington-transport-4pi}, and non-equilibrium quantum dynamics \cite{berges-cox,berges-aarts,berges-baier,berges-non,berges-aarts2,tranberg-smit,tranberg-aarts,tranberg-laurie}.
An advantage of the $n$PI method is that it provides a systemmatic expansion for which the truncation occurs at the level of the action. 
One major disadvantages is a violation of gauge invariance \cite{Smit2003,Zaraket2004}. 
A method to mimimize gauge dependence has been proposed \cite{sym-improvement}, and some issues with applying the method are discussed in \cite{Marko20156,Whittingham2016,Whittingham2017}.
Another significant difficulty with the $n$PI formalism is renormalization. The renormalization of the symmetric 2PI effective theory using a counterterm approach was developed by a number of authors over a period of several years \cite{vanHees2002,Blaizot2003,Serreau2005,Serreau2010}. 
The renormalization of non-symmetric theories is more subtle, but important for the study of phase transitions and Bose Einstein condensation. Phase transitions are studied in a scalar $\phi^4$ theory in \cite{Alford,Szep1,Szep2,Szep3,Szep4}, and in the $SU(N)$ Higgs model in 3 dimensions (where vertex divergences are absent) in \cite{Moore2014}. Bose Einstein condensation has been studied in Refs. \cite{Marko2014,Blaizot2017}.

All of the 2PI calculations cited above have used a counterterm method to perform the renormalization. 
The introduction of multiple sets of vertex counterterms is required, and the complexity of the procedure is such that it is unknown how to extend it to the 4PI theory.
However, since the introduction of higher order variational vertices is numerically very difficult, one might be tempted to ignore vertex corrections and try to improve previous 2PI calculations by increasing the order of the truncation (typically the loop order). 
In calculations where infrared divergences play an important role, such 2PI calculations at higher loop order can be useful \cite{sym-improvement,Szep4,Marko2014,Blaizot2017,mati-infraredqed}.
However, it is known that $n$PI formulations with $n>2$ are necessary in some situations. Transport coefficients in gauge theories cannot be calculated, even at leading order, using a 2PI formulation \cite{Carrington-transport-3pi}. Numerical calculations using a 
symmetric scalar $\phi^4$ theory have shown the importance of 4PI vertex corrections in 3 dimensions \cite{mikula}, and the breakdown of the 2PI approximation at the 4 loop level in 4 dimensions \cite{meggison}. 

There is a general hierarchial relationship between the order of the truncation and the number of variational vertices that can be included \cite{berges-hierarchy}.
If the effective action is truncated at $L$ loops in the skeleton expansion, the corresponding $n$PI effective actions are identical for $n\ge L$ 
(equivalently, one necessarily works with $L\ge n$). 
In this sense, a 3 loop calculation done within the 3PI formalism, a 4 loop calculation done within the 4PI formalism, etc, is complete. 
There is additional evidence that an $L$ loop calculation in the $n$PI formalism should, in general, be done with $L=n$.
In gauge theories, the $n$ loop $n$PI effective action respects gauge invariance to the order of the truncation \cite{Smit2003,Zaraket2004}, and one therefore expects that a 3 loop 2PI calculation will have stronger gauge dependence than a 3 loop 3PI one. 
In QED a 2 loop 2PI calculation (which is complete at 2 loop order according to the hierarchial relationship discussed above) found weak dependence on the gauge parameter \cite{Borsanyi-2pi}, while a recent 3 loop 2PI calculation in $SU(N)$ Higgs theory \cite{Moore2014} has found strong dependence on the gauge parameter. 

There is evidence therefore that higher order $n$PI calculations are important and worthwhile to persue.
Higher order effective actions have been derived using different methods \cite{berges-hierarchy,Carrington2004,Guo2011,Guo2012}, but little progress has been made in solving the resulting variational equations.
As mentioned above, one major problem is that the renormalization of such theories in 4 dimensions cannot be done (using any known method) by introducing counterterms. 
This is, in part, the reason that a method has been developed to apply the FRG to a 2PI theory 
\cite{Blaizot-2PIa,Blaizot-2PIb,Carrington-BS,pulver}.
Similar techniques have been used in a condensed matter context in \cite{Dupuis1,Dupuis2,Katanin}.

Another significant technical problem is the size of the phase space involved in self consistent calculations of vertex functions. 
Because of limitations of memory and computation time, very few calculations that include variational vertices have been done, and typically various ans\"atze are introduced for the vertex functions. 
In Ref. \cite{Moore2012}, the authors study Yang Mills QCD in the 3PI approximation, but they work in 3 dimensions and use Pad\'e ans\"atze for the variational functions. 
A set of self consistent vertex equations obtained from a 3 dimensional 3PI Yang Mills theory were solved in \cite{Huber2016}, but the full structure of the vertices was replaced with comparatively simple ans\"atze. 
Probably the most complete calculation to date was done in Ref. \cite{Fischer2016} where the authors study QCD at the 3 loop 3PI level. They use a clever technique to exploit the symmetry of the vertices and simplify the variational equations, but they do not actually solve the fully self consistent integral equations, but rather obtain the ghost and gluon propagators using a separate truncation, and input these results into the vertex calculations.

The ultimate goal of our research program is to do a 4 loop 4PI calculation. 
We work (so far) with a symmetric scalar theory, in order to avoid the complications associated with the Lorentz and Dirac structures of fields in gauge theories.
As described above, there are two main obstacles:
a conceptual one (renormalizability) and a technical one (memory and computation time contraints encountered because of the large phase space associated with self consistent vertices).
In this paper we develop a method to resolve both of these issues, and test it by performing a 4 loop 2PI calculation. 
We renormalize the theory using the FRG method that was introduced in \cite{pulver} at the 3 loop 2PI level. 
Using this method, no counterterms are introduced, and all divergences are absorbed into the bare parameters of the Lagrangian, the structure of which is fixed and completely independent of the order of the approximation. 
The RG method should therefore work at any loop order, and at any order in the $n$PI approximation. 
In \cite{pulver} we tested our RG method by applying it to a symmetric 3 loop 2PI calculation. 
However, at 3 loops the traditional calculation requires only one vertex counterterm, and in this sense does not involve the full complexity of the 2PI counterterm renormalization procedure. 
One could therefore suspect that the agreement of our 3 loop RG calculation with the standard counterterm calculation is an artifact of the approximation. 
One of the motivations for the calculation in this paper is to verify that this agreement holds at the 4 loop level.
The results of this paper show that a calculation that requires two vertex counterterms using the traditional method, can be done using the RG method by appropriately defining the one bare coupling constant that appears in the original Lagrangian. The success of this calculation is therefore strong evidence that the RG method will also work on the technically more difficult 4PI calculation. 

The success of our approach is verfied by comparing results with our previous calculation \cite{meggison} which
used Cartesian coordinates;
performed all integrals using fast Fourier transforms (which implement periodic boundary conditions);
and used counterterm renormalization. 
Cartesian coordinates are required when using fast Fourier transforms, but beyond the 2PI level they are impractical because of the size of the phase space involved when vertex functions are represented on a Cartesian grid. 
In order to reduce the size of the vertex phase space to a numerically manageable level, we use spherical coordinates.
However, switching to spherical coordinates means giving up the speed obtained from fast Fourier transforms. 
Adequate speed is obtained by exploiting the symmetries of the vertex function, and developing efficient interpolation and integration methods. 

This paper is organized as follows. In section \ref{section-preliminaries} we present our notation and the setup of the calculation. 
In section \ref{section-method} we describe our method and derive the flow equations that we will solve. In section \ref{section-numerical} we give some details of our numerical procedure. Our results are presented in section \ref{section-results}, and some further discussion and conclusions are give in section \ref{section-conclusions}.

\section{Preliminaries}
\label{section-preliminaries}
\subsection{Notation}
\label{section-notation}

In most equations in this paper we suppress the arguments that denote the space-time dependence of functions. As an example of this notation, the quadratic term in the action is written:
\bea
\frac{i}{2}\int d^4 x\,d^4 y\,\varphi(x)G_{\rm no\cdot int}^{-1}(x-y)\varphi(y) ~~\longrightarrow~~\frac{i}{2}\varphi\, G_{\rm no\cdot int}^{-1}\varphi\,.
\eea
The notation $G_{\rm no\cdot int}$ indicates the bare propagator.
The classical action is 
\bea
\label{action}
&& S[\varphi] =\frac{i}{2}\varphi \,G_{\rm no\cdot int}^{-1}\varphi -\frac{i}{4!}\lambda\varphi^{4}\,,~~
iG_{\rm no\cdot int}^{-1} = -(\Box + m^2)\,.
\eea
For notational convenience we use a scaled version of the physical coupling constant
($\lambda_{\,{\rm phys}} = i\lambda$). The extra factor of $i$ will be removed when rotating to Eucledian space to do numerical calculations.

Using the functional renormalization group method, 
we add to the action in (\ref{action}) a non-local regulator term
\bea
\label{action-RG}
S_{\kappa}[\varphi]=S[\varphi]+\Delta S_{\kappa}[\varphi]\,,~~~
\Delta  S_{\kappa}[\varphi] = -\frac{1}{2}\hat R_{\kappa}\varphi^2\,. 
\eea
The parameter $\kappa$ has dimensions of momentum. The regulator function is chosen to have the following
properties:  $\lim_{Q\ll \kappa}\hat R_{\kappa}(Q)\sim \kappa^{2}$ and $\lim_{Q\geq\kappa}\hat R_{\kappa}(Q)\rightarrow 0$. The effect is therefore that for $Q\ll \kappa$ the regulator acts like a large mass term 
which suppresses quantum fluctuations with wavelengths $1/Q\gg
1/\kappa$, but fluctuations with $Q\gg\kappa$ and wavelengths $1/Q\ll
1/\kappa$ are not affected by the presence of the regulator.

The $n$-point functions of the theory depend on the parameter $\kappa$.
One obtains a hierarchy of coupled differential `flow' equations for the derivatives of the $n$-point functions with respect to $\kappa$. 
This hierarchy is automatically truncated when the effective action is, and there is therefore no need to introduce additional approximations. 
The set of truncated flow equations can be integrated from an ultraviolet scale $\kappa=\mu$ down to $\kappa=0$ where the regulator goes to zero and the desired quantum $n$-point functions are obtained. 
The parameter $\mu$ is the scale at which the bare masses and couplings are defined (we use $\mu$ instead of the traditional $\Lambda$ because that letter will be used for a 4-point kernel).
One chooses $\mu$ large enough that when $\kappa=\mu$ the theory is classical. 
 The 2- and 4-point functions are then known functions of the bare parameters, and these classical solutions can be used as initial conditions on the differential flow equations. 

\subsection{The 2PI FRG effective action}
\label{section-action}

The 2PI generating functionals are calculated from the regulated action (\ref{action-RG}): 
\begin{eqnarray}
\label{ZandW-RG}
&& Z_\kappa[J,J_{2}] = \int[d\varphi]\exp\bigg\{i\Big(S[\varphi]+J\varphi
+\frac{1}{2}J_{2}\varphi^2 - \frac{1}{2}\hat R_\kappa\varphi^2\Big)\bigg\}\,,\\[2mm]
\label{ZandW-RG-2}
&& W_\kappa[J,J_{2}] = -i\ln Z_\kappa[J,J_{2}]\,,\\[2mm]
\label{expt-values-RG}
&& \frac{\delta W_\kappa[J,J_2]}{\delta J}
 = \langle \varphi\rangle\equiv\phi\,,~~~ \frac{\delta W_\kappa[J,J_2]}{\delta
J_{2}} = \frac{1}{2}\langle\varphi^2\rangle = \frac{1}{2}(\phi^2 + G)\,.
\end{eqnarray}
The 2PI effective action is obtained by taking the double Legendre transform of the generating functional $W_\kappa[J,J_2]$ with respect to the sources $J$ and $J_2$ and taking $\phi$ and $G$ as the independent variables:
\begin{eqnarray}
\label{legendre-transform-RG}
\hat\Gamma_\kappa[\phi,G]&=&W_\kappa-J\frac{\delta W_\kappa}{\delta J} -J_{2}\frac{\delta W_\kappa}{\delta J_{2}} = W_\kappa-J\phi-\frac{1}{2}J_{2}(\phi\phi+G)\,.
\end{eqnarray}
After performing the Legendre transform, the functional arguments of the effective action $\phi$ and $G$ are formally independent of the regulator function and the parameter $\kappa$, but the non-interacting propagator does depend on $\kappa$. 
We define 
\bea
\label{G0-first}
iG_{\mathrm{no}\cdot \mathrm{int}\cdot\kappa}^{-1} = iG_{\mathrm{no}\cdot \mathrm{int}}^{-1}-\hat R_\kappa = -\Box-(m^2+\hat R_k) \,.
\eea
Using this notation the effective action $\hat\Gamma_\kappa[\phi,G]$ can be written
\bea
\label{gamma-RG}
&& \hat\Gamma_\kappa[\phi,G] =\Gamma_{\mathrm{no}\cdot \mathrm{int}\cdot\kappa}[\phi,G] +\Gamma_{\mathrm{int}}[\phi,G]\,, \\[2mm]
&& \hat\Gamma_{\mathrm{no}\cdot \mathrm{int}\cdot\kappa}[\phi,G] = \frac{i}{2}\phi \,G_{\mathrm{no}\cdot \mathrm{int}\cdot\kappa}^{-1}\phi+\frac{i}{2}\mathrm{Tr}\ln
G^{-1}
+\frac{i}{2}\mathrm{Tr}G_{\mathrm{no}\cdot \mathrm{int}\cdot\kappa}^{-1}G \,,\nonumber\\[2mm]
&& \Gamma_{\mathrm{int}}[\phi,G] =  -\frac{i}{4!} \lambda \phi^{4}-\frac{i}{4}\lambda \phi^{2}G+\Gamma_{2}[\phi,G;\lambda]\,,\nonumber
\eea
where $\Gamma_{2}$ contains all 2PI graphs with two and more loops.
We define an effective action that corresponds to the original classical action at the scale $\mu$:
\bea
\label{hatGamma}
\Gamma_\kappa = \hat\Gamma_\kappa -\Delta S_\kappa(\phi)\,.
\eea

\begin{figure}[h]
\begin{center}
\includegraphics[width=12cm]{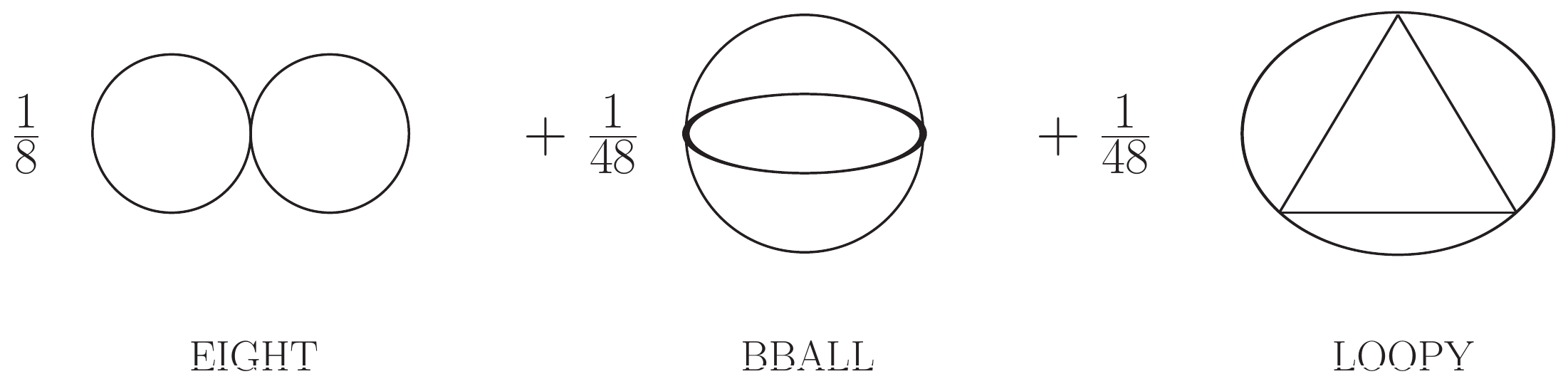}
\end{center}
\caption{The interacting part of the effective action for the symmetric theory to 4 loop order.  \label{phi-fig}}
\end{figure}

Throughout this paper we use the notation $\Gamma = -i \Phi$  where both $\Gamma$ and $\Phi$ carry the same subscripts or superscripts. For example, the 
interacting part of the action is written $\Gamma_{\rm int}[\phi,G] = -i \Phi_{\rm int}[\phi,G]$.
To make the equations look nicer we also define an imaginary regulator function $R_\kappa = -i \hat R_\kappa$ (the extra factor $i$ will be removed when we rotate to Eucledian space). Using this notation equations (\ref{G0-first} - \ref{hatGamma}) are rewritten
\bea
\label{phi-def}
&& \Phi_\kappa = \Phi_{\mathrm{no}\cdot \mathrm{int}\cdot\kappa}+ \Phi_{{\rm int}}\,,\\
\label{phi-def2}
&& \Phi_{\mathrm{no}\cdot \mathrm{int}\cdot \kappa} = -\big[\frac{1}{2}{\rm Tr}\,{\rm ln} G^{-1}+\frac{1}{2} G_{\mathrm{no}\cdot \mathrm{int}\cdot\kappa}^{-1}G\big]
- \big[\frac{1}{2}G_{\mathrm{no}\cdot \mathrm{int}\cdot\kappa}^{-1} + \frac{1}{2} R_\kappa\big]\phi^2\,,\\[2mm]
\label{G0-RG}
&& G_{\mathrm{no}\cdot \mathrm{int}\cdot\kappa}^{-1} = G_{\mathrm{no}\cdot \mathrm{int}}^{-1}-R_\kappa \,.
\eea
We  extremize the effective action by solving the variational equations of motion for the self consistent 1 and 2 point functions. These self consistent solutions will depend on the parameter $\kappa$ and are therefore denoted $\phi_\kappa$ and $G_\kappa$.
We will work throughout with the symmetric theory by setting $\phi_\kappa=0$. 
In figure \ref{phi-fig} we show the diagrams that contribute to $\Phi_{\rm int}$ to 4 loop order, and the names we will use for these diagrams. 

\subsection{Kernels}
\label{section-kernels}

\begin{figure}[htb]
\begin{center}
\includegraphics[width=14cm]{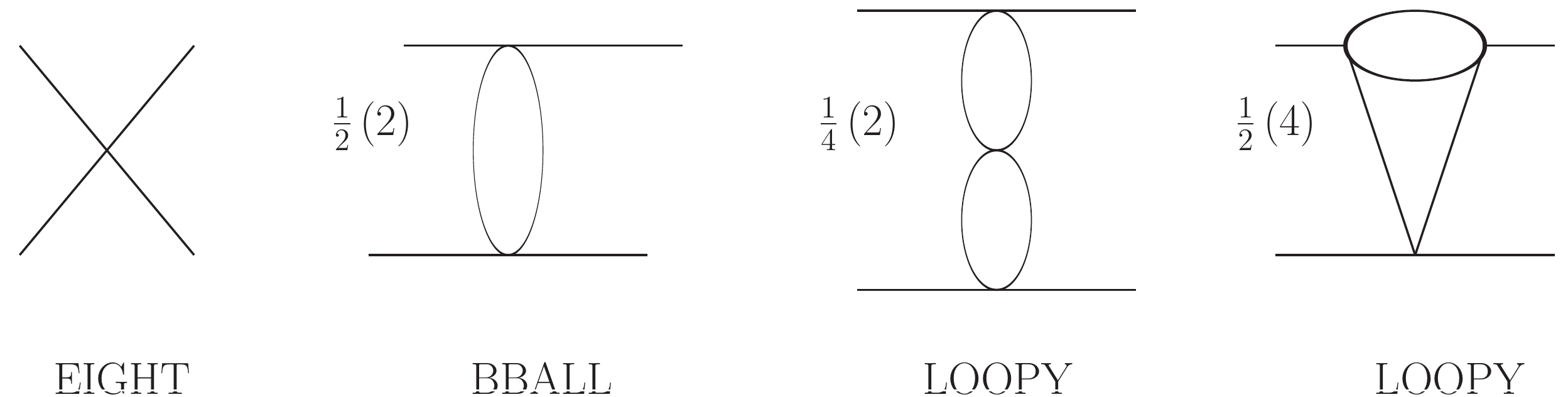}\\
\vspace*{1cm}
\includegraphics[width=13cm]{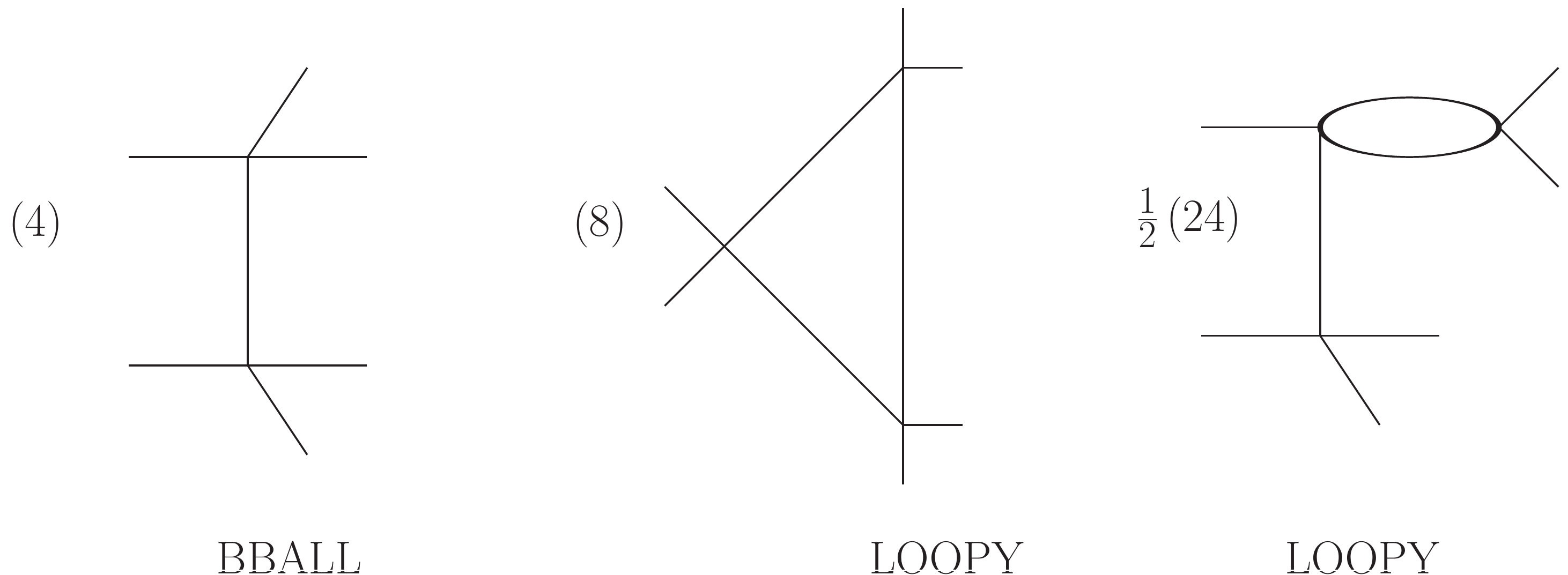}
\end{center}
\caption{Skeleton diagrams in the 4-kernel $\Lambda$ and the 6-kernel $\Upsilon$. The numbers in brackets indicate the number of distinct permutations of external legs, and the numbers that are not bracketed are symmetry factors. The name under each diagram indicates the term in the effective action that produced it (see Fig. \ref{phi-fig}). \label{lambda-fig}}
\end{figure}

We define a set of $n$-point kernels by taking functional derivatives of the effective action
\bea
\label{kernels-RG}
&& \Lambda^{(n,m)} = 2^m \frac{\delta^n}{\delta \phi^n}\frac{\delta^m}{\delta G^m} \Phi_{\mathrm{int}}\,.
\eea
Since we work with the symmetric theory, we consider only kernels with $n=0$ and we suppress the corresponding 0 index in the superscript. 
Substituting the self consistent solutions into the definition of the kernels (\ref{kernels-RG}) we obtain
\bea
\label{kernels-RG-kappa}
&& \Lambda_\kappa^{(m)} = \Lambda^{(m)} \bigg|_{\stackrel{G=G_\kappa}{\phi=o}}\,.
\eea
 We introduce specific names for the kernels we will write repeatedly:
\bea
\label{kernels-specific}
&& \Lambda^{(1)} = \Sigma\,,~~~\Lambda^{(2)} = \Lambda\,,~~~\Lambda^{(3)} = \Upsilon\,.
\eea
The kernels $\Sigma$, $\Lambda$ and $\Upsilon$ have two, four and six legs, respectively. 
The Fourier transformed functions are written $\Sigma(P)$, $\Lambda(P,K)$ and $\Upsilon(P,K,Q)$.
The diagrams that contribute to $\Lambda$ and $\Upsilon$ in the 4 loop approximation are shown in Fig. \ref{lambda-fig}. We note that the 4-kernel contains 2 loop diagrams that involve nastly overlapping divergences. 
When we use the RG method, any kernel that contains subdivergences is calculated from a flow equation. 
This is explained in detail in the next section.

\section{Method}
\label{section-method}

\subsection{Flow equations}
\label{section-flowequations}

Using the chain rule, and the fact that $\Phi_{\mathrm{int}}$ and therefore $\Lambda^{(m)}$ does not explicitly depend on $\kappa$, we find
\bea
\label{flow-eq}
 \partial_\kappa \Lambda^{(m)}_\kappa = \frac{1}{2}\,\partial_\kappa G_\kappa \,\Lambda_\kappa^{(m+1)}\,.
\eea
In momentum space (restoring arguments) this equation has the form
\bea
\label{flow-eq-mom}
&& \partial_\kappa\Lambda^{(m)}_\kappa(P_1,P_2,\cdots P_{m}) = \frac{1}{2}\int dQ \, \partial_\kappa G_\kappa(Q) \,\Lambda^{(m+1)}_\kappa(P_1,P_2,\cdots P_{m+1},Q)\,.
\label{flow-generic}
\eea
Equation (\ref{flow-eq-mom}) is an infinite heirarchy of coupled integral equations for the $n$-point kernels. 
This structure is typical of continuum non-perturbative methods, for example Schwinger-Dyson equations and traditional (1PI) RG calculations. In our formalism however, the hierarchy truncates at the level of the action. This can be seen immediately from equation (\ref{flow-eq-mom}) since it is clear that when the effective action is truncated at any order in the skeleton expansion, the kernel on the right side of (\ref{flow-eq-mom}) is zero when the largest number of propagators that appears in any diagram in the effective action is less than $m+1$. As will be explained in section \ref{section-consistency}, the hierarchy can be truncated at an even earlier level.

\begin{figure}[h]
\begin{center}
\includegraphics[width=14cm]{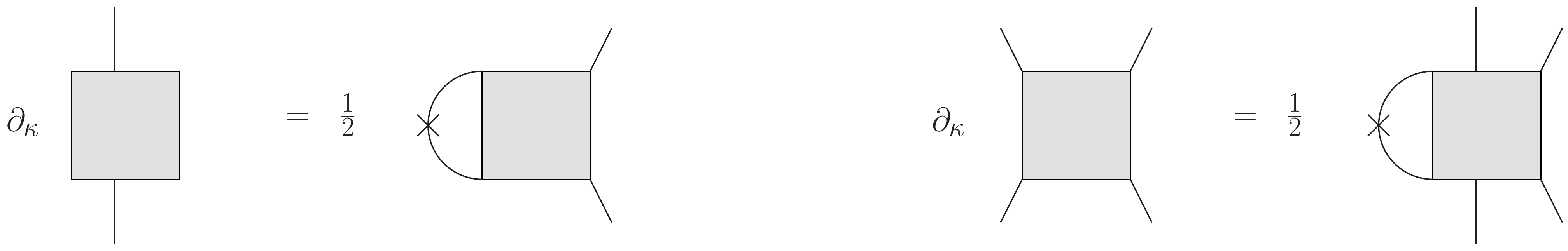}
\end{center}
\caption{The flow equations in equation (\ref{flow-sigma2}, \ref{flow-lambda22}).  \label{flow-fig}}
\end{figure}

We show below how to rewrite the flow equations (\ref{flow-eq-mom}) in a more convenient form. 
The stationary condition is 
\bea
\label{stat-cond}
\frac{\delta \Phi_\kappa[\phi,G]}{\delta G}\bigg|_{G=G_\kappa} = 0
\eea
and using equations (\ref{phi-def} - \ref{kernels-RG}, \ref{kernels-specific}) the variation produces a Dyson equation for the non-perturbative 2-point function in terms of the 2-kernel $\Sigma$:
\bea
\label{dyson1-RG}
&& G_\kappa^{-1}  =  G_{\mathrm{no}\cdot \mathrm{int}}^{-1} - R_\kappa - \Sigma_\kappa(\phi,G_\kappa)\,.
\eea
Using (\ref{dyson1-RG}) we have
\bea
\label{eq1a}
\partial_\kappa G_\kappa = -G_\kappa\,(\partial_\kappa G^{-1}_\kappa)\, G_\kappa
 = G_\kappa\,\big(\partial_\kappa(R_\kappa+\Sigma_\kappa)\big)\,G_\kappa\,.
\eea
The first two equations in the hierarchy (\ref{flow-generic}) can now be rewritten using (\ref{kernels-specific}, \ref{eq1a}) as
\bea
\label{flow-sigma2}
&& \partial_\kappa\Sigma_\kappa(P) = \frac{1}{2}\int dQ \, \partial_\kappa\big[\Sigma_\kappa(Q) + R_\kappa(Q) \big]\, G_\kappa^2(Q)\,\Lambda_\kappa(P,Q)\,,\\[.2cm]
\label{flow-lambda22}
&& \partial_\kappa \Lambda_\kappa(P,K) = \frac{1}{2}\int dQ \,\partial_\kappa \big[R_\kappa (Q)+\Sigma_\kappa (Q)\big]\, G^2_\kappa(Q) \,\Upsilon_\kappa(P,K,Q)\,.
\eea
These equations are shown in Fig. \ref{flow-fig}.
The grey boxes denote the kernels $\Sigma$, $\Lambda$ and $\Upsilon$ (the specific kernel is indicated by the number of legs attached to the box).
The crosses indicate the insertion $\partial_\kappa\big(\Sigma_\kappa(Q) + R_\kappa(Q)\big)$.

Finally we note that the flow equation for the 2-kernel $\Sigma$ can be rewritten in terms of the Bethe-Salpeter (BS) vertex. 
Iterating equation (\ref{flow-sigma2}) we obtain
\bea
\label{flow-sigma2-glob2}
\partial_\kappa \Sigma_\kappa(P) = \frac{1}{2}\int dQ \, \partial_\kappa R_\kappa(Q)\, G_\kappa^2(Q)\,M_\kappa(P,Q)\,,
\eea
with 
\bea
\label{bethe-sal}
M_\kappa(P,K) = \Lambda_\kappa(P,K)+\frac{1}{2}\int dQ \Lambda_\kappa(P,Q)\,G_\kappa^2(Q) M_\kappa(Q,K)\,.
\eea

In the 2PI formalism, we can also define non-perturbative vertices in terms of the change in the effective action with respect to variations in the field evaluated at the stationary point. These are usually called `physical' vertices. The physical 4 point function, which we call $V$, 
can be written in terms of the BS vertex as
\bea
\label{V-defn}
V = \lambda + 3\big(M - \Lambda) \,.
\eea
We note that the vertex $V$ involves a resummation in all three ($s$, $t$ and $u$) channels, but using our shorthand notation which suppresses indices, the three channels combine to produce the factor (3) in equation (\ref{V-defn}). Details of the derivation of (\ref{V-defn}) are given in Refs. \cite{Berges2005a,Carrington-BS}. 

In order to do numerical calculations, we rotate to Eucledian space by defining the Eucledian variables:
\bea
\label{euc-defn}
&& q_0 ~\to~ i q_4\,,~~dQ ~\to~i dQ_E\,,~~Q^2~\to ~ -Q_E^2 \,,\\
&& m^2 ~\to~ m^2_E\,,~~\lambda ~\to~ -i \lambda_E
\,,~~\delta \lambda ~\to~ -i\delta \lambda_E\,,\nonumber\\
&& G^{-1}_{\rm no\cdot int}~\to ~ i (G^{-1}_{\rm no\cdot int})_E\,,~~\Sigma~\to~ -i \Sigma_E  ~~\Rightarrow~~     G^{-1}~\to ~ i G_E^{-1}  = i(P_E^2+m_E^2+\Sigma_E)\,,\nonumber \\
&& \Lambda ~\to~ i\Lambda_E\,,~~\Upsilon_\kappa \to -i \Upsilon_{\kappa E} \,,~~M ~\to~ i M_E\,,\,,~~V ~\to~ i V_E\,.\nonumber\\
&& R_\kappa=-i R_{\kappa\,E}\,.
\eea
The extra factors of $i$ in the definitions of $\lambda_E$ and $R_E$ remove the factors that were introduced in the definitions $\lambda_{\rm phys}=i\lambda$ (under equation (\ref{action})) and $\hat R=i R$ (under equation (\ref{hatGamma})).  
From this point forward we suppress the subscripts which indicate Eucledian space quantities. 
The flow equations (\ref{flow-sigma2}, \ref{flow-lambda22}) and the BS equation (\ref{bethe-sal}) have the same form in Eucledian space.
The Eucledian space Dyson equation is
\bea
 G^{-1}(P) = G_{\rm{no\cdot int}}^{-1}(P) +\Sigma(P)\,,
\eea 
and the physical vertex becomes
\bea
V = -\lambda + 3\big(M - \Lambda) \,.
\eea
The regulator function we use has the Eucledian momentum space form
\bea
\label{Rdef}
R_\kappa(Q) = \frac{Q^2}{e^{Q^2/\kappa^2}-1}\,.
\eea

\subsection{Tuning}
\label{section-tuning}
Physical considerations dictate the renormalization conditions which are enforced on the non-perturbative quantum $n$-point functions that are obtained at the $\kappa=0$ end of the flow. We use standard renormalization conditions:
\bea
\label{rc-euc}
G_0^{-1}(0) = m^2\,,~~~M_0(0,0) = -\lambda\,.
\eea
The quantum $n$-point functions in (\ref{rc-euc}) are obtained by solving the integro-differential flow equations (\ref{flow-sigma2}, \ref{flow-lambda22}), starting from some set of initial conditions at $\kappa=\mu$.
The regulator function $R_\kappa$ (equation (\ref{Rdef})) is chosen so that at the scale $\kappa=\mu$ the theory is described by the classical action and the initial conditions for the flow equations can be taken from the bare masses and couplings of the original Lagrangian. 

It is clear that one must know the values of the bare parameters from which to start the flow at the beginning of the calculation. 
A different choice of bare parameters will  give different quantum $n$-point functions at the end of the flow, and therefore different renormalized masses and couplings. 
The procedure to figure out the values of the bare parameters that will satisfy the chosen renormalization conditions is called tuning. 
Starting from an initial guess for the bare parameters, we solve the flow equations, extract the renormalized parameters, adjust the bare parameters either up or down depending on the result, and solve the flow equations again. 
The calculation is repeated until the chosen renormaliation condition is satisfied to the desired accuracy.

\subsection{Consistency}
\label{section-consistency}

Since RG procedure involves initializing the flow equations at an ultraviolet scale $\kappa=\mu$, and also enforcing renormalization conditions at the $\kappa=0$ end of the flow, we must address the question of whether or not the initial and renormalization conditions can be defined consistently. 

Consider an arbitrary $n$-point kernel of the form
\bea
\label{n-generic}
\Lambda^{(m)}_\kappa(P_1,P_2\dots ) = \tilde\Lambda^{(m)}_\kappa(P_1,P_2\dots ) + C(P_1,P_2\dots)
\eea
where $\tilde\Lambda^{(m)}(P_1,P_2\dots)$ is the result obtained from integrating the corresponding flow equation, and $C(P_1,P_2\dots)$ is a $\kappa$ independent integration constant. 
In the limit $\kappa\to\mu$ we require that the vertex function approaches a momentum independent constant:
\bea
\label{mulimit}
\lim_{\kappa\to\mu} \Lambda^{(m)}_\kappa(P_1,P_2\dots ) = \Lambda^{(m)}_\mu(P_1,P_2\dots ) \equiv -\lambda_\mu
\eea
Comparing equations (\ref{n-generic}, \ref{mulimit}) we have
\bea
\label{Cdef}
C(P_1,P_2\dots) = -\lambda_\mu - \tilde\Lambda^{(m)}_\mu(P_1,P_2\dots)\,,
\eea
so that (\ref{n-generic}) becomes
\bea
\label{n-generic-2}
\Lambda^{(m)}_\kappa(P_1,P_2\dots ) = -\lambda_\mu + \tilde\Lambda^{(m)}_\kappa(P_1,P_2\dots ) -\tilde\Lambda^{(m)}_\mu(P_1,P_2\dots ) \,.
\eea

Now we look at the $\kappa\to 0$ end of the flow and determine the conditions under which we can enforce the renormalization condition $\Lambda_0(0,0,\dots) = -\lambda$. 
We start by adding and subtracting two different terms to the original vertex, and grouping into square brackets the differences we will consider below. This gives
\bea
\label{differences}
&& \Lambda^{(m)}_\kappa(P_1,P_2\dots)  \\
&& = -\lambda + \big[\Lambda^{(m)}_\kappa(P_1,P_2\dots) - \Lambda^{(m)}_0(P_1,P_2\dots) \big] 
+ \big[\Lambda^{(m)}_0(P_1,P_2\dots) - \Lambda^{(m)}_0(0,0,\dots) \big] \,.\nonumber
\eea
The renormalization condition is satisfied if the square brackets go to zero in the limit that $\kappa$ and the momentum arguments go to zero. 
Setting $\kappa=0$ and using equation (\ref{n-generic-2}) we obtain
\bea
\big[~~\big] = \big(\tilde\Lambda^{(m)}_0(P_1,P_2\dots) - \tilde\Lambda^{(m)}_0(0,0\dots)\big) - \big(\tilde\Lambda^{(m)}_\mu(P_1,P_2\dots) - \tilde\Lambda^{(m)}_\mu(0,0\dots)\big)\,.
\label{second-difference}
\eea
The second term in (\ref{second-difference}) is zero in the limit $\mu \gg P$, since this is (by construction) the limit in which loop contributions are suppressed and the momentum dependence of the vertex disappears. 
We have therefore shown that the renormalization condition will be consistent with the initial condition if
\bea
\label{final-condition}
{\cal Z} = \lim_{(P_1,\;P_2\;\dots)\to 0} \big(\tilde\Lambda^{(m)}_0(P_1,P_2\dots) - \tilde\Lambda^{(m)}_0(0,0\dots)\big) ~~\to~~ 0\,.
\eea
In the next section we will show that this condition is  satisfied if the truncation of the hierarchy in (\ref{flow-eq-mom}) is performed correctly.


\vspace*{.5cm}

We comment on the fact that the discussion above is misleading in one important way. 
It appears that the bare vertex $\lambda_\mu$ cancels exactly when we go from equation (\ref{n-generic-2}) to equation (\ref{differences}). 
If this were true, the initial condition and the renormalization condition would be unconnected to each other, and tuning would not be possible. The apparent cancellation occurs because the self consistent nature of the set of coupled equations for  kernels with different numbers of legs is not evident when we consider one kernel in isolation. 
%

Finally, we note that renormalization conditions are typically defined on vertices constructed from resummed kernels (see equation (\ref{rc-euc})) and not the kernels themselves. 
However, the condition derived above (\ref{final-condition}) is still sufficient to guarantee the consistency of the procedure. 
The crucial point is that the kernels are 2-particle-irreducible (see Fig. \ref{lambda-fig}) and therefore when they are chained together to form a resummed BS vertex, no additional subdivergences are produced. 
It is easy to see how this works in our calculation, where we only need to enforce a renormalization condition on the 4 point function (as will be explained in the next section). When we define the renormalization condition on the BS vertex instead of the kernel $\Lambda$, the result is only a shift in the final value of the bare coupling $\lambda_\mu$ that is produced by the tuning procedure. 

\subsection{Truncation}
\label{section-truncation}

Now we return to the issue of the truncation of the hierarchy of flow equations in equation (\ref{flow-eq-mom}).
%
%
The kernels obtained from direct functional differentiation of the action using (\ref{kernels-RG-kappa}) will automatically satisfy the correct flow equations (\ref{flow-eq-mom}). As explained in section \ref{section-flowequations}, this is just an obvious application of the chain rule. 
Next we observe that if a given kernel obtained from functional differentiation satisfies the condition (\ref{final-condition}), then using the analysis of the previous section we know it will also satisfy $\Lambda^{(m)}_0(0,0\cdots)=-\lambda$ and $\Lambda^{(m)}_\mu(0,0\cdots)=-\lambda_\mu$. The conclusion is that we do not need to solve its flow equation (since the result from solving the flow equation would be precisely equal to the expression obtained from the functional integration, with the addition of the appropriate constant).
The smallest value of $m$ for which (\ref{final-condition}) is satisfied is the `terminal' kernel which truncates the hierarchy of flow equations. 
After we have identified the terminal kernel, the set of flow equations for the kernels with $2\times(1,2,3,\dots m-1)$ legs can then be solved self consistently. 

The final step is to show that the flow equation for each kernel can be initialized at the classical solution, which is just the corresponding bare coupling.
This is just a consequence of the structure of the flow equations, in which the kernel with $2m$ legs is constructed by calculating a 1 loop integral obtained by joining two legs of the kernel with $2(m+1)$ legs (see figure \ref{flow-fig}). If $\Lambda_0^{(m+1)}$ is finite up to a momentum independent bare coupling constant, then clearly the result for $\Lambda_0^{(m)}$ obtained from solving a flow equation of the form (\ref{flow-eq-mom}) will also be finite. \\

In order to identify the terminal kernel, we need to know under what circumstances the condition (\ref{final-condition}) will be satisfied by a given 2PI kernel $\Lambda^{(m)}_\kappa$.
We start by looking at an example where it will not. We consider the self energy diagram on the right side of Figure \ref{nested-fig} which gives
\bea
\Sigma_0(P)=
 -\frac{\lambda^2}{6}\int dQ\int dL \,G_0(L)G_0(L+Q)G_0(P+Q)\,.
\eea
This sunset contribution to the self energy is produced by the BBALL diagram in the effective action (see Fig. \ref{phi-fig}). 
The quantity ${\cal Z}$ in equation (\ref{final-condition}) now takes the form
\bea
{\cal Z} && = -\frac{\lambda^2}{6} \int d Q\int d L \,  G_0(L) G_0( L+ Q) \big[ G_0( P + Q)- G_0( Q)\big]\,, \\
&& = -\frac{\lambda^2}{6} P^2 \int d Q\int d L \, G_0(L) G_0( L+ Q) \big[ G_0^\prime( Q) + \cdots\big]\,,
\eea
where in the last line we have expanded around $P^2=0$. 
The prime denotes differentiation with respect to $ Q^2$ and the dots represent terms that are higher order in $P^2/Q^2$.
It is clear that the divergent $L$ integral is unaffected by the subtraction, and therefore we cannot conclude that ${\cal Z}\to 0$ when $P$ approaches zero. 
%

Another example is the contribution to the 4-kernel $\Lambda$ from the last diagram in the first line of Fig. \ref{lambda-fig}. If we route the momenta as shown in the right side of Fig. \ref{nested-fig} and rescale momenta as described above, the divergence in the $L_2$ integration is unaffected by the subtraction, and therefore the condition (\ref{final-condition}) is not satisfied. 

In general, any loop that does not necessarily carry one of the external momenta is a `bad' loop. 
If a kernel does not have any bad loops, it satisfies (\ref{final-condition}) and its flow equation does not have to be solved.
The smallest of these kernels is the terminal kernel. 
In the following two subsections we explain how to apply these ideas to the 2PI calculation at the 3 and 4 loop level. 

\subsubsection{3 loop}

If we truncate the effective action at the 3 loop (BBALL) level, the self energy includes the sunset diagram which has a bad loop, as explained above. The kernel $\Lambda$ has no bad loops,  since the 1 loop BBALL contributions in Fig. \ref{lambda-fig} always carry external momenta. The vertex $\Lambda$ is therefore the terminal kernel and can be substituted directly into the $\Sigma$ flow equation.
In order to satisfy the initial condition, we replace the tree vertex (the EIGHT contribution in Fig. \ref{lambda-fig}) with the bare parameter $-\lambda_\mu$. 
The $\Lambda$ flow equation (\ref{flow-lambda22}) is not affected by any constant shift of the 4-kernel. 
Thus we see that we can obtain, directly from the action,  an expression for the 4-kernel that obeys the initial condition and satisfies the correct flow equation. 
We have also explicitly checked that the results are the same as those obtained from  solving the coupled set of $\Sigma$ and $\Lambda$ flow equations. 

\subsubsection{4 loop}
\begin{figure}[h]
\begin{center}
\includegraphics[width=12cm]{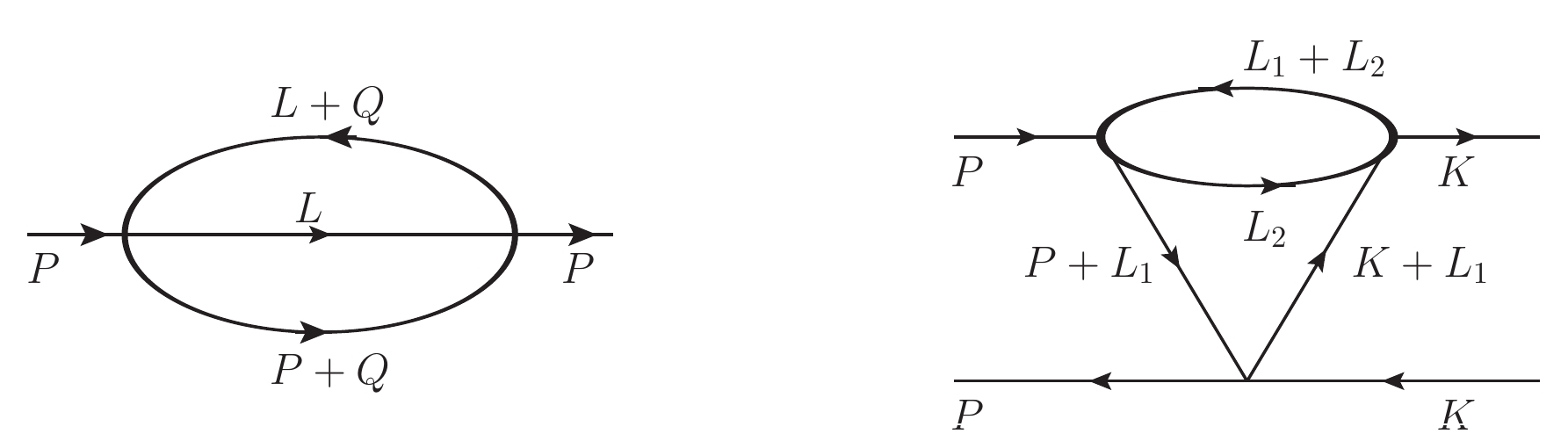}
\end{center}
\caption{Example graphs.  \label{nested-fig}}
\end{figure}

At the 4 loop level, the kernel $\Lambda$ contains a bad loop (as illustrated in the right side of Fig. \ref{nested-fig}) and does not satisfy (\ref{final-condition}). We therefore {\it cannot} not use the explicit expression for the 4 kernel $\Lambda$ shown in Fig. \ref{lambda-fig} directly in the flow equation for the 2 kernel (equation (\ref{flow-sigma2})), as we did at the 3 loop level. 
The kernel $\Upsilon$ (the bottom line of Fig. \ref{lambda-fig}) contains only 1 loop diagrams that always carry external momenta, and therefore satisfies (\ref{final-condition}) and is the terminal kernel. 
Since there is no bare 6-vertex in the Lagrangian, we know the integration constant should be set to zero. 
We therefore substitute the result for $\Upsilon$ shown in Fig. \ref{lambda-fig} directly into the $\Lambda$ flow equation. 
The $\Sigma$ and $\Lambda$ flow equations must then be solved self consistently.

\section{Numerical Method}
\label{section-numerical}

\subsection{Procedure}
\label{section-procedure}

We initialize the flow of the 2 and 4 kernels
\bea
\label{startG}
&& \Sigma_\mu(P) = m_\mu^2-m^2 \,,\\
\label{startL}
&& \Lambda_\mu(P,K) = -\lambda_\mu\,,
\eea
and we take the propagator in the ultraviolet limit as $G_\mu^{-1}(P) = P^2+m_\mu^2$.

Schematically the numerical procedure can be described as follows:

\begin{enumerate}
\item Choose values for the physical mass $m$ and coupling $\lambda$. We use always $m=1$, which means we give all dimensionful quantities in mass units. 

\item Calculate the quantum $n$-point functions:

\begin{enumerate}
\item Initialize the differential flow equations using (\ref{startG}, \ref{startL}).

\item Guess at the correct value for the bare parameters $m_\mu$ and $\lambda_\mu$.

\item Substitute the result for the 6-kernel $\Upsilon$ obtained from the 4 loop 2PI effective action (shown in the second part of Fig. \ref{lambda-fig}) into the $\Lambda$ flow equation.

\item Solve the integro-differential flow equations:
\bea 
&& \Sigma_\mu(P) ~~\longrightarrow ~~ \Sigma_0(P)\quad \text{and}\quad G_0 = \big(P^2+m^2+\Sigma_0(P)\big)^{-1} \,,\\
&& \Lambda_\mu(P,K) ~~\longrightarrow ~~ \Lambda_0(P,K)\,.
\eea

\item Solve the BS equation
\bea
M_0 = \Lambda_0+\frac{1}{2} \Lambda_0 G_0^2 M_0\,.
\eea

\item Extract renormalized mass and coupling 
\bea
&& m^2_{\rm found} = G_0^{-1}(0) = m^2+\Sigma_0(0)\,,\\
&& -\lambda_{\rm found} = M_0(0,0)\,.
\eea
\end{enumerate}

\item Compare the chosen and found values for the mass and coupling, adjust the bare values up or down accordingly, and repeat all steps until the renormalization conditions are satisfied to the desired accuracy. 
\end{enumerate}

\subsection{Parameters}
\label{section-parameters}

The differential equations are solved using a logarithmic scale by defining the variable $t=\ln \kappa/\mu$, in order to increase sensitivity to the small $\kappa$ region where we approach the quantum theory. We use $\kappa_{\rm max} = \mu=100$, $\kappa_{\rm min}=10^{-2}$ and $N_\kappa=50$. We have checked that our results are insensitive to these choices. In addition, we have checked for possible dependence on the form of the regulator function by using a generalization of (\ref{Rdef}):
\bea
\label{Rdef-gen}
R_\kappa(Q;z) = \frac{\kappa^2 (Q^2/\kappa^2)^z}{(e^{Q^2/\kappa^2}-1)^z}\,.
\eea
The original expression (\ref{Rdef}) corresponds to the choice of exponent $z=1$. 
Using $z=1/2$ and $z=1/4$ produces results that are virtually identical. \\

The 4-dimensional momentum integrals are written in the imaginary time formalism as
\bea
\label{4dint}
\int dK \,f(k_0,\vec k) = \sum_n \int\frac{d^3k}{(2\pi)^3}\,f(m_t n,\vec k)\,,
\eea 
with $m_t = 2\pi T$. 
Numerically we take $N_t$ terms in the summation with $\beta = \frac{1}{T} = N_t a_t$ where the parameter $a_t$ is the lattice spacing in the temporal direction. 

The integrals over the 3-momenta are done in spherical coordinates and using Gauss-Legendre integration. 
We use typically $N_x=N_\phi=8$ points for the integrations over the cosine of the polar angle and the azimuthal angle. 
The dependence on these angles is weak, and results are very stable when $N_x$ and/or $N_\phi$ are increased. 
To calculate the integral over the magnitude of the 3-momenta, we define a spatial length scale analogous to the inverse temperature $L=a_s N_s$ where $a_s$ is the spatial lattice spacing and $N_s$ is the number of lattice points.

\subsection{Restrictions}
\label{section-restrictions}
The numerical method replaces a continuous integration variable with infinite limits by a discrete sum over a finite number of terms. 
For numerical accuracy, we need generically that the upper limit of the sum is big and the step size is small. This means we require
$p_{\rm max}\sim \frac{1}{a_s} \gg 1$ and $\Delta p \sim \frac{1}{L}=\frac{1}{N_s a_s} \ll 1$.
The number of lattice points $N_s$ is limited by memory and computation time, and therefore there is a limit on how small $a_s$ can be taken while maintaining $N_s a_s$ large. 
However, there is another more subtle issue that limits how small we can choose $a_s$.
The theory has a Landau pole at a scale that decreases when $\lambda$ increases. 
When $\lambda$ becomes large, $a_s$ must increase ($p_{\rm max}$ must decrease) so that the integrals are cut off in the ultraviolet at a scale below the Landau scale. 
However, decreasing the ultraviolet cutoff $p_{\rm max}$ will eventually cause important contributions from the momentum phase space to be missed. 
When $\lambda$ has increased to the point that the Landau scale has moved down and dipped into the momentum regime over which the integrand is large, physically meaningful results cannot be obtained. 
In our calculation we have determined that the maximum coupling we can calculate is $\lambda\approx 5$. 

Finally, we note that it is well known that  scalar $\phi^4$ theory in 4-dimensions is non-interacting if it is considered
as a fundamental theory valid for arbitrarily high momentum scales (quantum triviality), but the renormalized coupling is non-zero if the theory has an ultraviolet cutoff and an infrared regulator. In our calculation the mass $m$ regulates the infrared and the lattice spacing parameter provides an ultraviolet cutoff. 

\section{Results and Discussion}
\label{section-results}

\begin{figure}[htb]
\begin{minipage}{12.5cm}
\center
\includegraphics[width=1.02\textwidth]{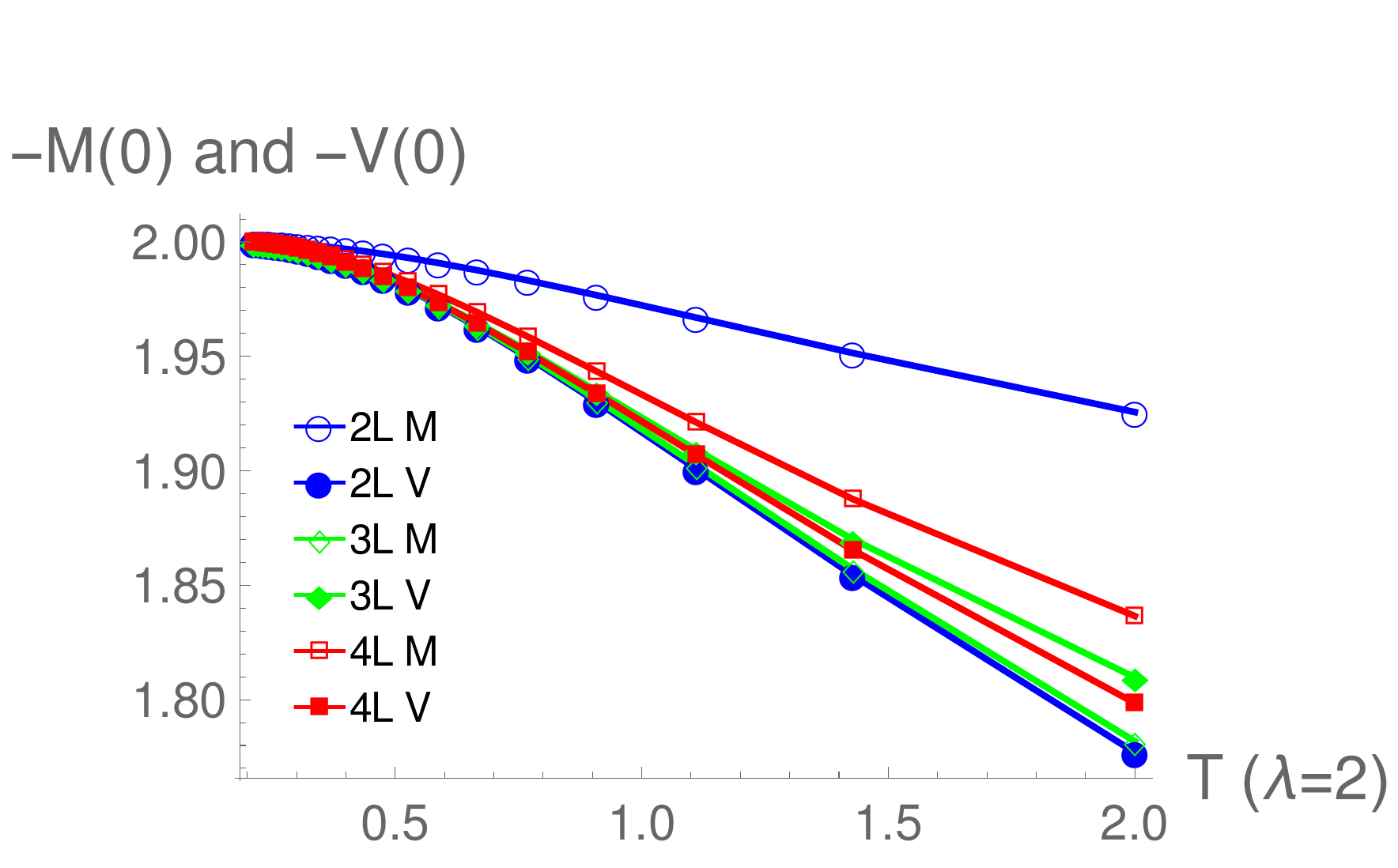}
\end{minipage}
\begin{minipage}{12.5cm}
\center
\includegraphics[width=1.02\textwidth]{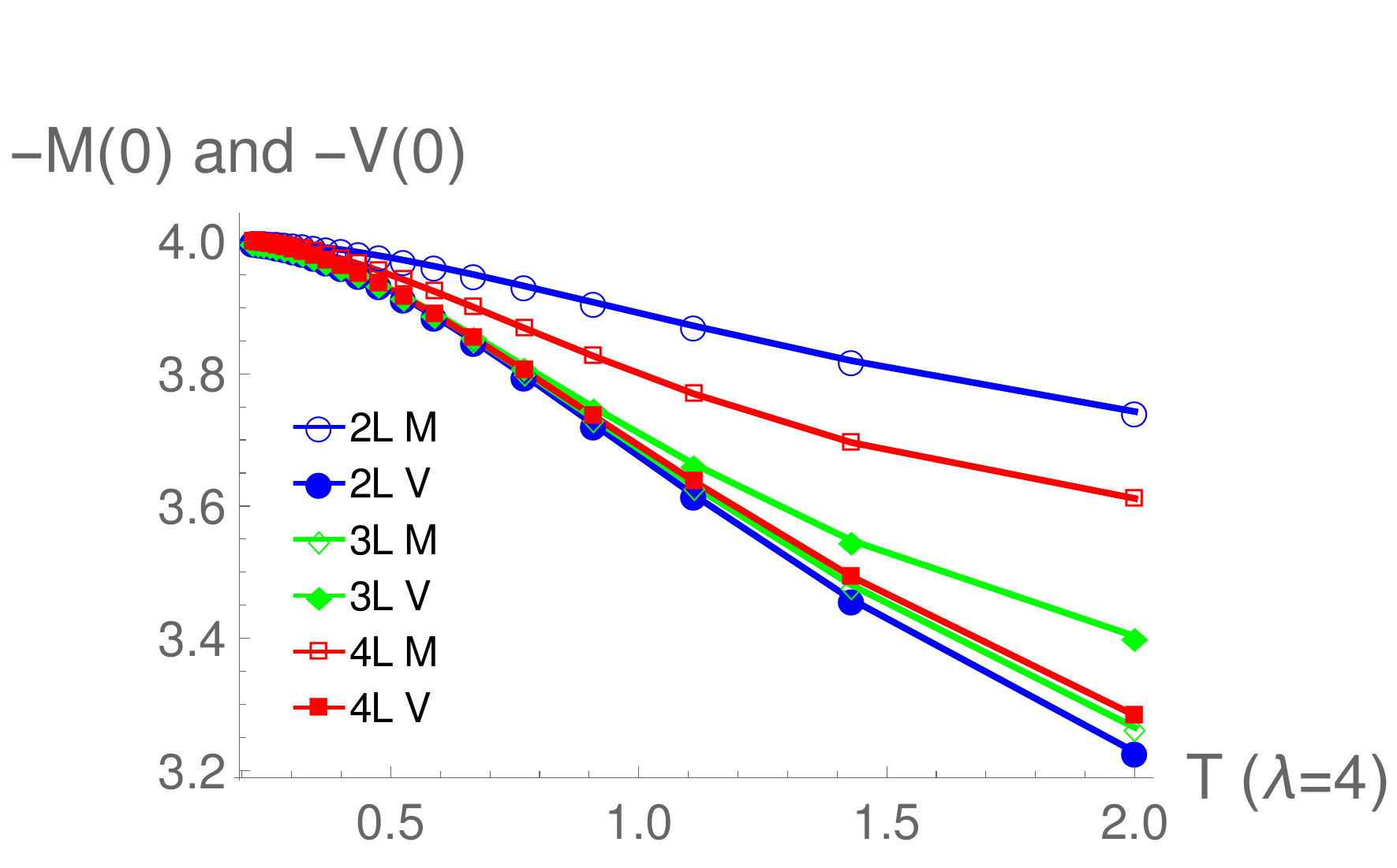}
\end{minipage}
\caption{The Bethe-Salpeter vertex $M(0)$ and the physical vertex $V(0)$ versus $T$ for $\lambda=2$ and $\lambda=4$ at the 2, 3 and 4 loop levels in the skeleton expansion of the action.}
\label{MVversusT}
\end{figure}

We use $a_t=1/10$, $a_s=1/5$ and $N_s=18$ and vary the temperature by changing the number of lattice points in the temporal direction. The renormalization is done at $N_t=37$ 
and the highest temperature we consider is $T=2$ which corresponds to $N_t=5$.
In Fig. \ref{MVversusT} we show the BS and physical vertices as functions of the temperature. The graph agrees well with the results of our previous calculation \cite{meggison} which used two separately defined counterterms. Some differences between the two calculations are expected, due to the different boundary conditions that must be implemented using the spherical and cartesian/fft methods. 

\begin{figure}[htb]
\center
\includegraphics[width=0.65\textwidth]{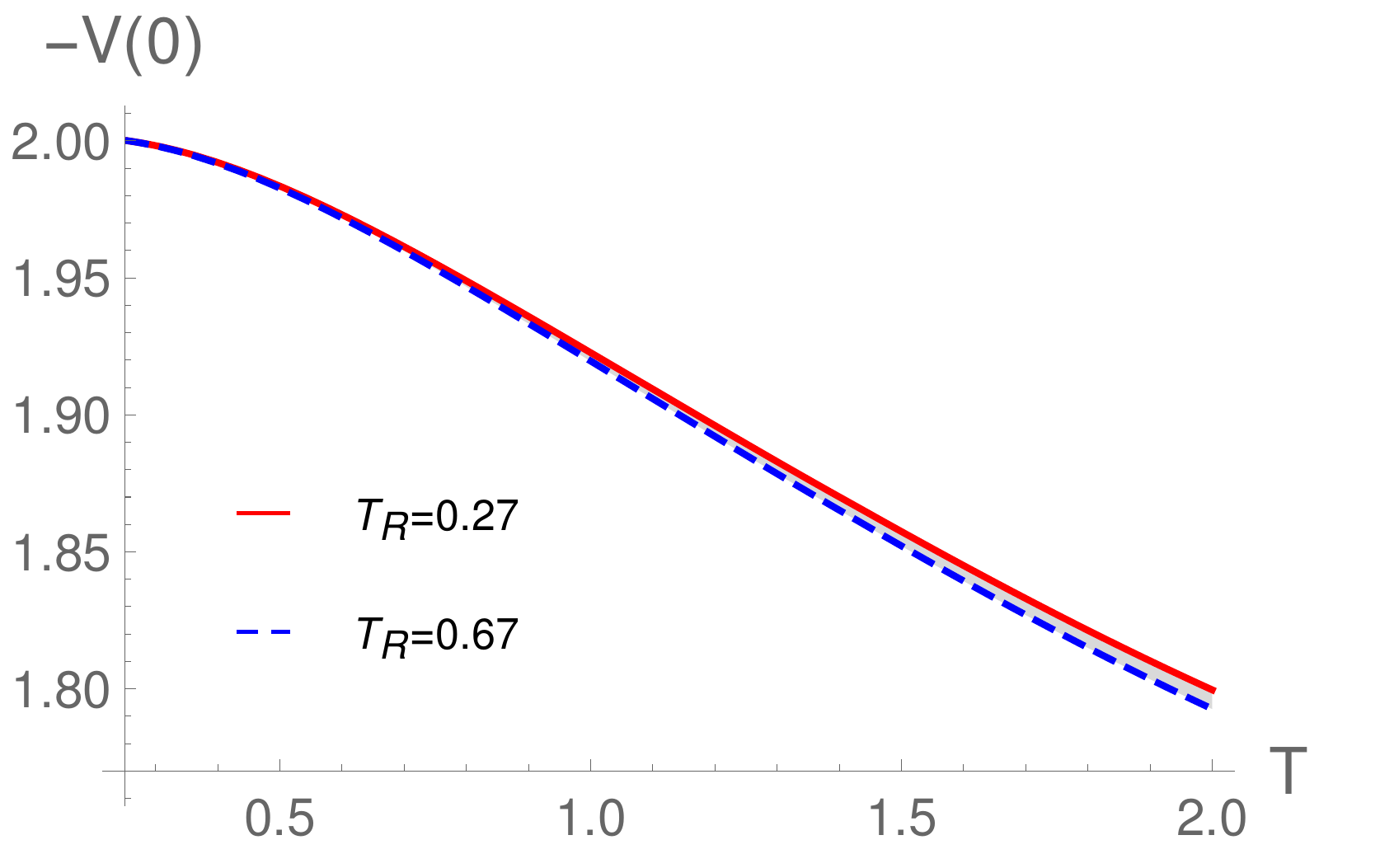}
\caption{The physical vertex $V(0)$ versus $T$ for $\lambda=2$ with the renormalization performed at two different temperatures. }
\label{Vscale}
\end{figure}

\begin{figure}[htb]
\center
\includegraphics[width=0.65\textwidth]{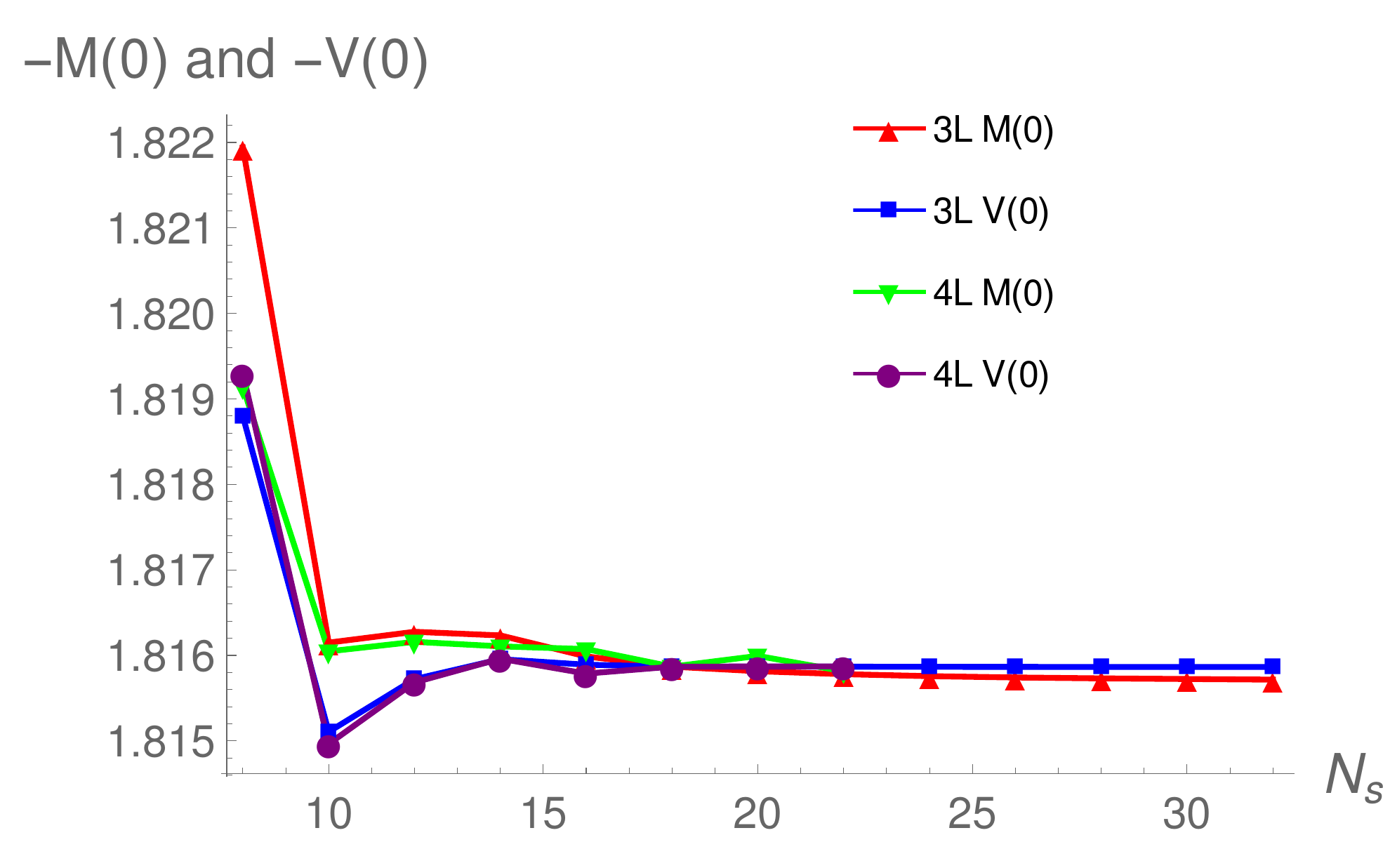}
\caption{The Bethe-Salpeter vertex $M(0)$ and the physical vertex $V(0)$ versus the number of spatial lattice points $N_S$ for $\lambda=T=2$ at the 3 and 4 loop levels in the skeleton expansion of the action. The curves are shifted so that they cross at $N_s=18$.}
\label{Lvariations}
\end{figure}

In Fig. \ref{Vscale} we show the dependence on the renormalization scale. The two curves correspond to the physical vertex $V(0)$ versus temperature with the renormalization done at two different temperatures. The scale dependence of the calculation is indicated by the shaded grey region between the two curves, and is very small.
In Fig. \ref{Lvariations} we show the BS and physical vertices versus $N_s$ with $a_s=1/5$ and $\lambda =T=2$. The grid spacing in momentum space is $\Delta p \sim \frac{1}{L} = \frac{5}{N_s}$. The graph shows that results are stable with $N_s\gtrsim 14$. The results in figures \ref{MVversusT} and \ref{Vscale} are produced with $N_s=18$, and the curves in this figure \ref{Lvariations} are shifted so that they cross at $N_s=18$, in order to provide the best means of comparison.  

\begin{figure}[htb]
\center
\includegraphics[width=0.60\textwidth]{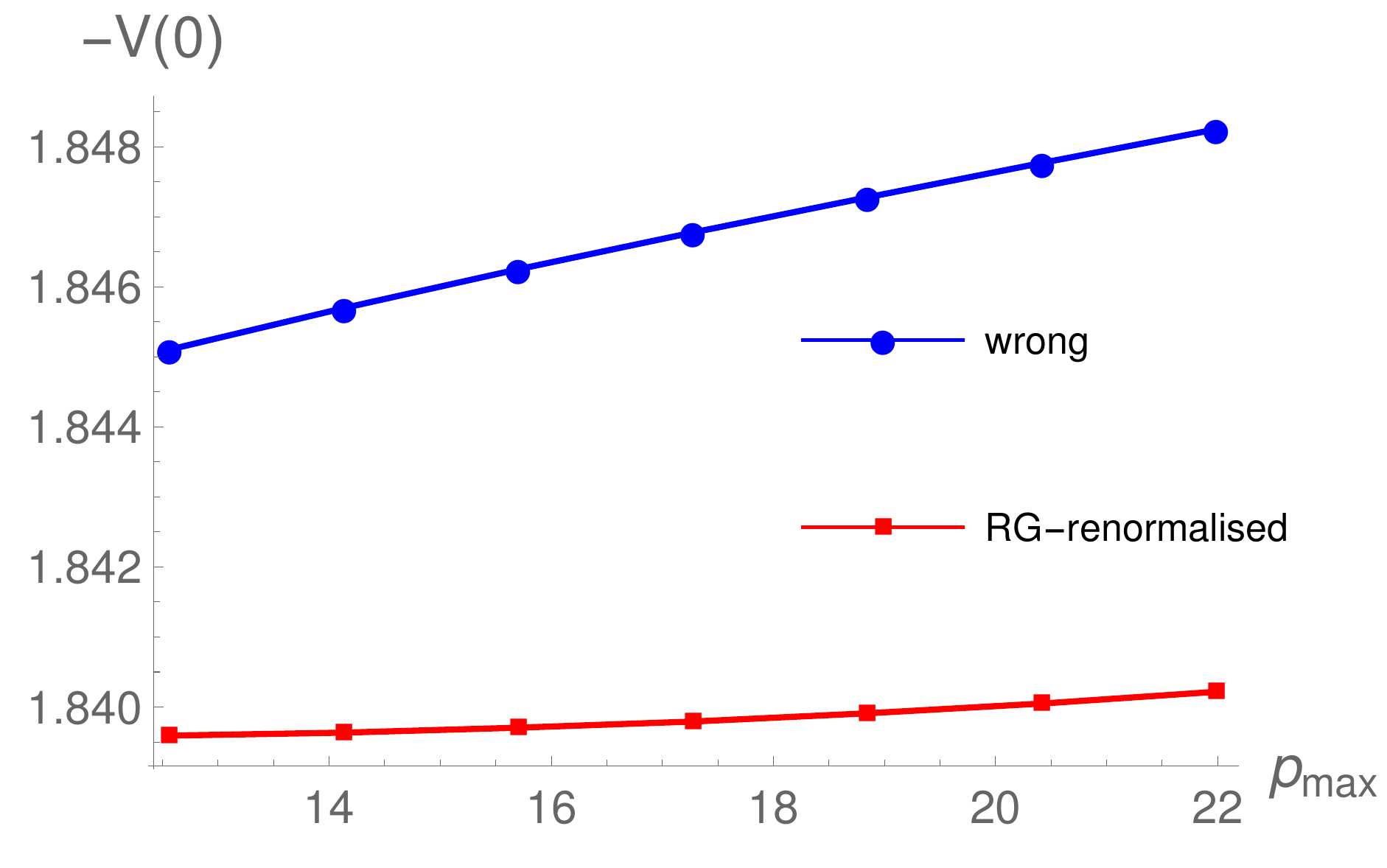}
\caption{The physical vertex $V(0)$ versus $p_{\rm max}$ with $\lambda=T=2$ and $L=4$ at the 4 loop level in the skeleton expansion of the action. For comparison, the results of an incorrect calculation are shown (see the text for further explanation). }
\label{as-data}
\end{figure}

In order to test the renormalization, we check that the results are unchanged when $p_{\rm max}$ is increased while $\Delta p\sim 1/L$ is held fixed. This is done by increasing $N_s$ while adjusting $a_s$ so that $L=a_s N_s$ is constant. In Fig. \ref{as-data} we use $L=4$ and $\lambda=T=2$. To set the scale, we compare with an incorrect 3 loop calculation, in which  one of the vertices in the 4 kernel is replaced with a bare vertex. 
At $p_{\rm max}\gtrsim 20$ the influence of the Landau pole is seen.
The data in figures \ref{MVversusT} and \ref{Vscale} is produced with $a_s=1/5$ or $p_{\rm max} = 15.71$.

\section{Conclusions}
\label{section-conclusions}

In this paper we have done a 4 loop 2PI calculation in a symmetric $\phi^4$ theory. 
We have renormalized the theory using the FRG method that was introduced in \cite{pulver} at the 3 loop 2PI level. 
Using this method, no counterterms are introduced, and all divergences are absorbed into the bare parameters of the Lagrangian, the structure of which is fixed and completely independent of the order of the approximation. 
We therefore expect that our RG method will work at any order in the $n$PI approximation. 
The next step in our program is to apply our method to a 4 loop 4PI calculation. 
The structures of the flow and Bethe-Salpeter equations are the same \cite{Russell2013}, but there is now a variational 4 vertex that must be calculated self-consistently. In spherical coordinates, the phase space for this vertex is comparable with that of the 3 dimensional self consistent 4PI vertex function that was calculated in \cite{mikula}. This calculation is currently in progress.

\end{document}